\documentclass[aps,prb,preprint,superscriptaddress,onecolumn,showpacs,amsmath,floatfix,citeautoscript]{revtex4}
\usepackage{graphicx}
\usepackage{float}
\usepackage{dcolumn}
\usepackage{color}
\usepackage{latexsym,bm}
\usepackage[normalem]{ulem}
\usepackage{multirow}
\usepackage{appendix}
\usepackage{epstopdf}

\newcommand{\MC}[1]{\textcolor{black}{{#1}}}

\begin{document}

\hyphenpenalty=5000
\tolerance=1000

\title{Retention and Recycling of Deuterium in Liquid Lithium-Tin Slab Studied by First-Principles Molecular Dynamics}

\author{Daye Zheng}
\affiliation{Key Laboratory of Quantum Information, University of Science and
  Technology of China, Hefei, Anhui, 230026, People's Republic of China}
\affiliation{Synergetic Innovation Center of Quantum Information and Quantum
  Physics, University of Science and Technology of China, Hefei, 230026, People's Republic of China}
\author{Zhen-Xiong Shen}
\affiliation{Key Laboratory of Quantum Information, University of Science and
  Technology of China, Hefei, Anhui, 230026, People's Republic of China}
\affiliation{Synergetic Innovation Center of Quantum Information and Quantum
  Physics, University of Science and Technology of China, Hefei, 230026, People's Republic of China}
\author{Mohan Chen}
\email{mohanchen@pku.edu.cn}
\affiliation{CAPT, HEDPS, College of Engineering, Peking University, Beijing 100871, People's Republic of China}
\author{Xinguo Ren}
\affiliation{Key Laboratory of Quantum Information, University of Science and
  Technology of China, Hefei, Anhui, 230026,  People's Republic of China}
\affiliation{Synergetic Innovation Center of Quantum Information and Quantum
  Physics, University of Science and Technology of China, Hefei, 230026,
  People's Republic of China}
\author{Lixin He}
\email{helx@ustc.edu.cn}
\affiliation{Key Laboratory of Quantum Information, University of Science and
  Technology of China, Hefei, Anhui, 230026,  People's Republic of China}
\affiliation{Synergetic Innovation Center of Quantum Information and Quantum
  Physics, University of Science and Technology of China, Hefei, 230026, People's Republic of China}
\date{\today}
%%%%%%%%%%%%%%%%%%%%%%%%%%%%%%%%%%%%%%%%%%%%%%%%%%%%%%%%%%%%%%%%%%%%%%%%%%%%%%%%%%%%%%%%%%%%%%%%%%%%%%%%%%%%%%%%%%%%%%%%%%%%%%%%%%%
%%%%%     Title
%%%%%%%%%%%%%%%%%%%%%%%%%%%%%%%%%%%%%%%%%%%%%%%%%%%%%%%%%%%%%%%%%%%%%%%%%%%%%%%%%%%%%%%%%%%%%%%%%%%%%%%%%%%%%%%%%%%%%%%%%%%%%%%%%%%

\begin{abstract}
{
Understanding the retention and recycling of hydrogen isotopes in liquid metal plasma-facing materials
such as liquid Li, Sn, and Li-Sn are of fundamental importance in designing magnetically confined fusion reactors.
We perform first-principles molecules dynamics simulations of liquid Li-Sn slab with
inserted D atoms to provide microscopic insights into the interactions of D with Li-Sn liquid metal.
\MC{We prepare two samples with low and high concentrations of D atoms.}
We observe evaporation of D$_2$ molecules out of the Li-Sn slabs \MC{in both concentrations of D}.
With detailed analysis, we unveil a cooperative process of forming D$_2$ molecules in liquid Li-Sn,
where Li atoms act as catalytic centers to trap a D atom before another D comes nearby to form a molecule, and the surplus
charges are transferred from D$_2$ to nearby Sn atoms.
%The above mechanism reduces the retention of D in the liquid Li-Sn slab, and is beneficial
%for recycling of D atoms.
Furthermore, we predict a temperature window \MC{in the low concentration case} in which D$_2$ molecules can escape to vacuum, while LiD molecules cannot.
The above findings deepen our understanding of interactions between hydrogen isotopes and Li-Sn liquid metal.
}

%Keywords: Liquid lithium-tin, plasma-facing components, molecular dynamics.
\end{abstract}
%\pacs{}
\maketitle

\section{Introduction}
Plasma-facing components (PFCs) in magnetic confinement fusion devices
are designed to withstand high heat loads of bombardments from energetic particles.
Meanwhile, hydrogen (H) isotopes such as deuterium (D) and tritium (T) are utilized as fuels
and their interactions with PFCs need to be thoroughly understood \cite{02JNM-Causey}.
In particular, the retention and recycling of D/T attract many attentions
due to their impacts on the fueling and vacuum pumping systems,
and also the plasma conditions \cite{00FED-Mattas}.
Here, {\it retention} and {\it recycling} respectively
refer to the hydrogen isotopes that remain in PFCs and return back to plasma
after the PFCs receive particle bombardments.
On one hand, a high retention rate of hydrogen isotopes suggests
that the temperature of core plasma is less affected by
preventing cold atoms from reentering the plasma, therefore,
the plasma performance could be increased with PFCs that
can retain a large amount of D/T atoms \cite{13PP-Majeski,15JNM-Abrams,15NF-Abrams}.
On the other hand, a high recycling rate of H isotopes can significantly reduce the fuel cost,
and a low retention of D/T atoms in PFCs also reduces safety concerns \cite{00FED-Mattas}.

The emergence of liquid-metal PFCs~\cite{14PS-Coenen} began receiving widespread attention in recent years.
Unlike solid metallic PFCs that unavoidably suffer from
mechanical failure problems caused by high energetic particles~\cite{12NF-Lipschultz, 13JNM-Pitts, 14NF-Eden},
liquid metals for PFCs own a low melting point,
resist erosion and neutron damage to a large extent, and has
high heat dissipation capabilities.
For instance, lithium (Li) has a low melting point of 453 K and
was found to improve the plasma parameters in a variety of experiments~\cite{13PP-Majeski,14FED-Fiflis,14PPCF-Tritz,
15JNM-Abrams,15NF-Abrams,17FED-Ono, 17FED-Maingi, 17L-Boyle, 18NF-Rindt}.
Tin (Sn) has also been tested in several experiments~\cite{15JNM-Morgan,16L-Eden,18NME-Cremona}
due to its low melting point of 505 K,~\cite{12JAP-Weir}
and additionally owns a low vapor pressure that is beneficial for high operating temperatures.
Nevertheless, plasma contamination due to accumulation of impurities atoms in plasma
is concerned since Sn has a high atomic number.
In this regard, Li-Sn eutectic is considered as an alternative candidate for PFCs with
combined advantages of both Li and Sn. However, the Li-Sn alloy
has been investigated in a few experiments \cite{04FED-Bastasz,04FED-Allain,07B-Allain,17NME-Kvon,17NME-Loureiro}.
Therefore, it is of great help to utilize first-principles computational methods
to provide fundamental insights into understanding liquid metals~\cite{13MP-Chen,18JCP-Beatriz,19NME-Beatriz}
and their interactions with hydrogen isotopes~\cite{13L-Krstic,16NF-Chen,17JCP-Liu,19NF-Beatriz}.

Experiments have found that hydrogen isotopes interact differently
with liquid Li and liquid Sn. For example,
it is known that hydrogen isotopes can be largely absorbed by liquid Li
due to their strong chemical affinity with Li atoms~\cite{92JNM-Moriyama,05JNM-Fukada,17FED-Kul,17FED-Taz}.
In contrast, the low reactivity of Sn with hydrogen isotopes leads to
a small retention ratio of hydrogen isotopes in liquid Sn~\cite{18NME-Cremona}.
It is intuitively expected that D atoms can be trapped by Li atoms in liquid Li-Sn eutectic,
resulting in a higher number of retained hydrogen isotopes in liquid Li-Sn than liquid Sn.
Surprisingly, a recent experiment~\cite{17NME-Loureiro} observed that Li-Sn eutectic
has a smaller retention ratio of D than pure Sn
against intuition.
Although the above experiment may suffer from contaminants such as oxygen and carbon at the surface of the sample,
it is still worth understanding
the fundamental processes regarding the interactions of hydrogen isotopes with liquid Li-Sn. For example,
how do D atoms escape from Li-Sn eutectic, causing the low retention ratio of D in Li-Sn?
More importantly, what is the underlying mechanism that explain
liquid Li, Sn, and Li-Sn interact differently with D?

In this work, we performed first-principles molecular dynamics (FPMD) simulations
to tackle the above issues.
We observe the formation and evaporation of D$_2$ molecules
in the Li-Sn slab, which may be a reason for the low retention ratio of D observed in experiment~\cite{17NME-Loureiro}.
These results are remarkable, because as predicted by recent FPMD studies,
D$_2$ molecules hardly form in liquid Li because D atoms
form strong ionic bonds with Li atom \cite{16NF-Chen},
whereas forming D$_2$ molecules is also difficult in liquid Sn,
which may be caused by the low reactivity of D in liquid Sn \cite{17JCP-Liu}.
In addition, it is still inconclusive which molecules can escape the surfaces of Li-Sn slab.

Very recently, an independent work~\cite{19NF-Beatriz} has also found the formation of D$_2$ molecules
and D$_2$ bubbles in bulk liquid Li-Sn systems with FPMD simulations.
However, the detailed forming processes of D$_2$ molecules have not been fully explored.
Furthermore, due to the lack of vacuum in the study, the evaporation of molecules was not investigated.
Herein, \MC{we prepare two Li-Sn slabs with low and high concentrations of D atoms.}
We unveil a cooperative mechanism of forming D$_2$ molecules in liquid Li-Sn slabs,
where Li atoms play a central role to trap a D atom, and wait for a second D
to come nearby, triggering the forming process of D$_2$ molecule.
During the formation of D$_2$ molecule,
the surplus electrons are transferred from D$_2$ to its adjacent Sn atoms.
D$_2$ molecules then diffuse rapidly by leaving their adjacent Li and Sn atoms.
Importantly, \MC{in the low concentration case,}
we find the existence of both D$_2$ molecules and LiD molecules in the vacuum area of the simulated cell.
Our results show that LiD molecules are more difficult to form than D$_2$ molecules \MC{below 873 K},
and we predict a optimal temperature window \MC{in the low concentration case} that allows evaporation of D$_2$ molecules
from surfaces of PFCs but not for LiD molecules,
which may potentially contaminate the plasma.
The above findings offer a series of new microscopic insights into our understanding
of how hydrogen isotopes interact with liquid Li-Sn eutectic.

\section{Computational Details}
All of the FPMD simulations were performed with
the ABACUS (Atomic-orbital Based Ab-initio Computation at USTC) package~\cite{16CMS-Li}.
We adopted norm-conserving pseudopotentials \cite{81PRB-Perdew}
and local density approximation \cite{80PRL-Ceperley}.
We chose an energy cutoff of 120 Ry for charge density and only the gamma point to sample the Brillouin zone with periodic boundary conditions.
We used numerical atomic orbitals~\cite{10JPCM-Chen} as basis sets to construct the electronic wave functions. Specifically,
double-zeta plus polarized (DZP) orbital sets were adopted for D ($2s1p$), Li ($2s1p$) and Sn ($2s2p1d$)~\cite{10JPCM-Chen, 11JPCM-Chen}.
The radius cutoffs of numerical atomic orbitals were chosen to be 6.0, 9.0 and 8.0 bohr for D, Li, and Sn elements, respectively.
Born-Oppenheimer molecular dynamics simulations were performed in the canonical ensemble NVT
%(constant number of particles $N$, constant volume $V$, and constant temperature $T$)
with the Nos\'{e}-Hoover thermostat~\cite{84JCP-Nose, 85PRA-Hoover} at 573, 673, and 873 K.
A liquid Li-Sn slab with a vacuum and two concentrations of D atoms
in a periodic cell ($60\times 14.98 \times 14.98$ \AA$^3$ where the vacuum is placed along $x$-axis) was studied.
A Li$_{0.19}$Sn$_{0.76}$D$_{0.05}$ slab containing 36 Li, 144 Sn, and 10 D atoms
is referred as the low concentration system,
while a Li$_{0.13}$Sn$_{0.51}$D$_{0.36}$ slab consisting of
36 Li, 144 Sn, and 100 D atoms is referred as the high concentration system for brevity.
\MC{We performed FPMD simulations with 30 ps trajectories for the Li$_{0.19}$Sn$_{0.76}$D$_{0.05}$ slab
at temperatures of 573, 673, and 873 K.
In addition, we ran FPMD simulations with 20, 24, and 22 ps trajectories for
the Li$_{0.13}$Sn$_{0.51}$D$_{0.36}$ slab at temperatures of 573, 673, and 873 K, respectively.
For comparison, we also carried out FPMD simulations for bulk Li$_{0.2}$Sn$_{0.8}$
(36  Li and 144 Sn atoms) and Li$_{0.19}$Sn$_{0.76}$D$_{0.05}$ (36 Li, 144 Sn, and 10 D atoms) systems.
The ionic densities of the two bulk systems were obtained by
performing FPMD simulations in the NVT ensemble with varying cell volumes
until the averaged pressure is less than 1 GPa;
the resulting cell sizes of the bulk Li$_{0.2}$Sn$_{0.8}$ and
Li$_{0.19}$Sn$_{0.76}$D$_{0.05}$ systems are 23.14$\times$14.20$\times$14.20~\AA$^3$
and 23.01$\times$14.12$\times$14.12~\AA$^3$, respectively.
We generated 20 ps trajectories for the two systems.
The time step of FPMD simulations utilized in systems with D atoms was set to 0.2 fs
to better characterize the D atoms that have a small mass; for those
 Li-Sn systems without D atoms, the time step was set to 0.5 fs.}

The liquid configurations of Li-Sn slabs were generated by performing FPMD simulations as follows.
\MC{
We first ran FPMD simulations of a Li$_{0.20}$Sn$_{0.80}$ slab for 5.0 ps at 1073 K with a time step of 0.5 fs in order
to obtain its liquid structure.
Next, we randomly inserted 10 and 100 deuterium (D) atoms into the Li$_{0.20}$Sn$_{0.80}$ system
to generate two systems with different D concentrations, i.e.,
Li$_{0.19}$Sn$_{0.76}$D$_{0.05}$ and Li$_{0.13}$Sn$_{0.51}$D$_{0.36}$ systems, respectively.
We then ran FPMD simulations at selected temperatures, i.e., 573, 673 and 873 K.
The lengths of the trajectories have been described above.
In order to let the systems evolve towards the equilibrium state,
our analysis was based on the last 20 ps trajectories.}

\section{Results and Discussions}

\begin{figure}
  \centering
  \includegraphics[width=0.6\textwidth]{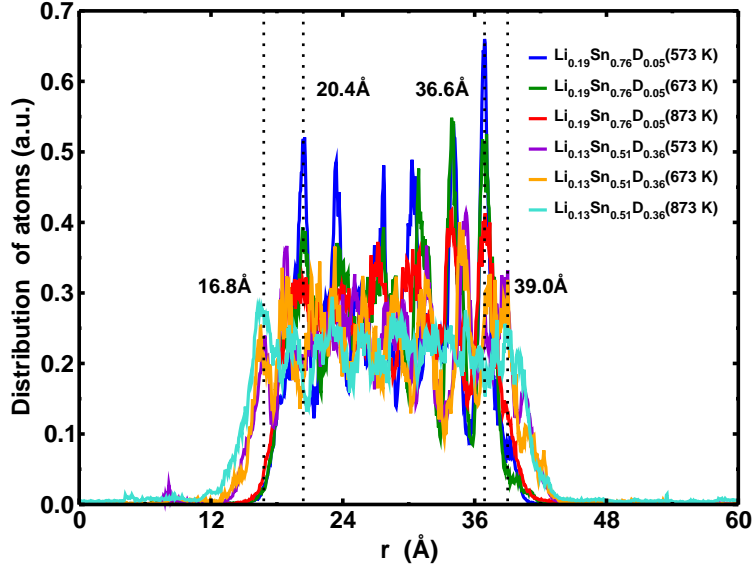}\\
  \caption{(Color online) Ionic density profiles of
  the Li$_{0.19}$Sn$_{0.76}$D$_{0.05}$ and Li$_{0.13}$Sn$_{0.51}$D$_{0.36}$ slabs
  along the surface normal direction at 573, 673, and 873 K.
  The two dashed lines at 20.4 and 36.6 \AA~represent the surfaces of the Li$_{0.19}$Sn$_{0.76}$D$_{0.05}$ Li-Sn slab,
  while the other two dashed lines at 16.8 and 39.0 \AA~indicate the surfaces of the Li$_{0.13}$Sn$_{0.51}$D$_{0.36}$ Li-Sn slab.
 }\label{fig:surface}
\end{figure}

\begin{figure}
  \centering
  \includegraphics[width=0.6\textwidth]{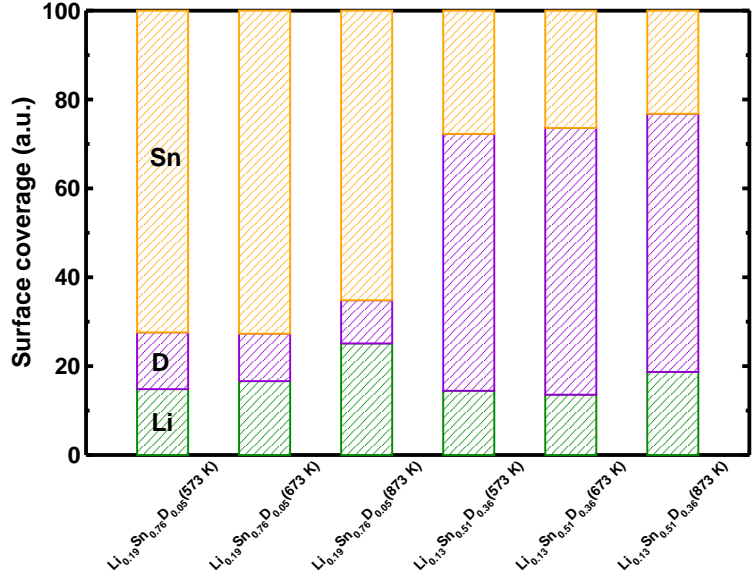}\\
  \caption{(Color online) Surface coverages of Li, Sn, and D atoms in the Li$_{0.19}$Sn$_{0.76}$D$_{0.05}$ and
  Li$_{0.13}$Sn$_{0.51}$D$_{0.36}$ slabs at 573, 673, and 873 K.}\label{fig:surface2}
\end{figure}

\begin{figure}
  \centering
  \includegraphics[width=0.7\textwidth]{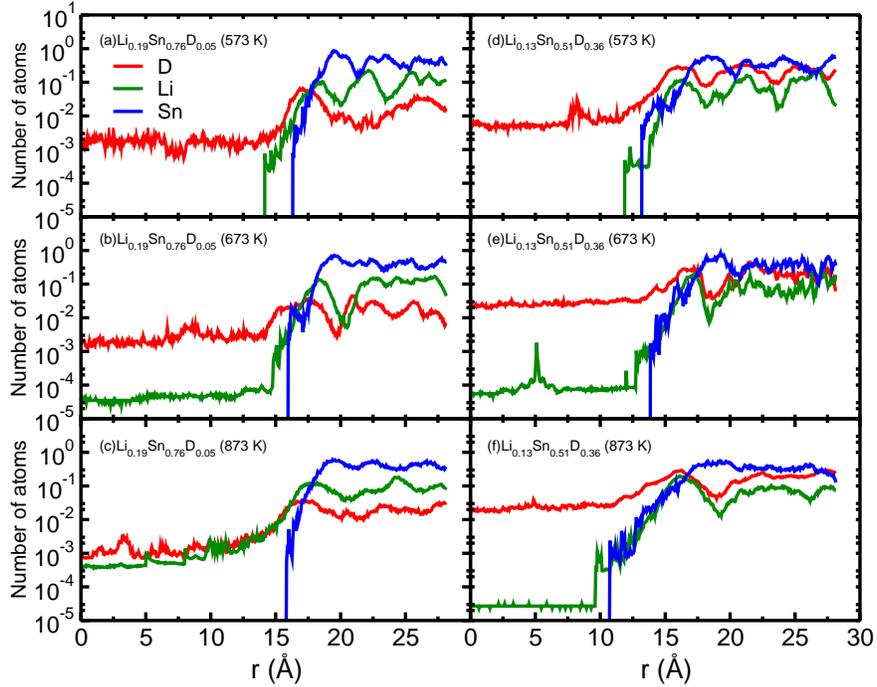}\\
  \caption{\MC{(Color online) Counted number of atoms per snapshot of D, Li, and Sn atoms along the surface normal direction $r$
  of the Li$_{0.19}$Sn$_{0.76}$D$_{0.05}$ and Li$_{0.13}$Sn$_{0.51}$D$_{0.36}$ slab systems.
  The density profiles are averaged over the two surfaces. The interval of $r$ is chosen to be 0.06~\AA. }}\label{fig:decompose}
\end{figure}

\subsection{Li-Sn Surfaces}
% define the slab
Figure~\ref{fig:surface} illustrates the simulated ionic density profiles of liquid Li-Sn systems
with both low and high concentrations of D atoms at 573, 673, and 873 K.
\MC{
We can see that the volume of a Li-Sn slab is insensitive to temperatures ranging from 573 to 873 K.
This is due to the fact that Li-Sn eutectic is a metallic liquid,
whose volume increases by about 1\% per 100 degrees.
For temperature that increases from 573 to 873 K, the length of the slab
along the surface normal direction ($x$ direction) slightly changes.
Therefore, we define the surface of a Li-Sn slab
as the space between the first peak of the ionic density profile
and the vacuum as shown in Figure~\ref{fig:surface},
while the bulk of the slab is defined as the space between the two surfaces of the slab.
Since the volume of a Li-Sn-D slab is insensitive to temperatures,
we take the Li-Sn-D slab systems at 673 K to define the bulk region of the slab.
Specifically, we define the bulk region of the Li$_{0.19}$Sn$_{0.76}$D$_{0.05}$ slab system as
the space ranging from $x$=20.4 to 36.6~\AA~as shown in Figure 1.
In addition, the volume of the Li-Sn-D slab substantially increases with a higher concentration of D.
In this regard, the bulk region of the Li$_{0.13}$Sn$_{0.51}$D$_{0.36}$ slab system is defined as
the space covering from $x$=16.8 to 39.0~\AA.
}
We also observe that the ionic density profiles of Li-Sn systems exhibit a series of sharp peaks,
\MC{which may be due to the finite size effects. However,}
the heights of peaks decrease at higher temperatures,
suggesting a more uniform distribution of atoms in the slabs at higher temperatures.
\MC{We also compute the mean square displacements (MSD) of Li and Sn atoms
in both low- and high-concentration systems at 573 K
and found the increase of MSD with respect to time (see Supplementary Materials), suggesting that atoms
are in the liquid state.
}

\MC{
Figure~\ref{fig:surface2} shows the surface compositions of Li-Sn-D slabs from simulations.
For both Li-Sn-D systems, we observe a similar trend as compared to experiment~\cite{04FED-Bastasz,04FED-Allain,07B-Allain} that
the surface coverage of Li increases at a higher temperature,
which can be explained as Li reduces the surface tension more effectively than Sn~\cite{04FED-Bastasz,17NME-Kvon}.
In addition, the surface coverage of D only slightly decreases with increasing temperature;
however,} in the high-concentration Li-Sn system,
the surface coverage of D is twice more than that of Sn,
although the total number of Sn atoms is slightly larger than the number of D atoms.

\MC{Figure~\ref{fig:decompose} illustrates the density profiles of Li-Sn-D slab systems decomposed
into atomic components.
Both low- and high-concentration slab systems at temperatures of 573, 673, and 873 K are considered.
We average two surfaces of a slab in order to yield more converged results.
We find that the outmost surface areas are dominated by Li atoms rather than Sn atoms
in all of the six systems shown in Figure~\ref{fig:decompose}, which
suggests that Li atoms are segregated to the surface of the Li-Sn-D slab system.
}

\MC{
However, since the FPMD simulations are extremely expensive,
%with conditions of a small time step (0.2 fs)
%and a large size of vacuum (about 30~\AA~along the surface normal direction),
the lengths of our FPMD trajectories are limited to a few tens of ps
and the system only consists of a few hundred atoms.
It would be possible that the segregation of Li to the surface is faster in a short time scale because
the velocities of Li atoms are higher and the system has not yet reached the equilibrium state.
Additionally, the sharp peaks also imply that the trajectory length may not be long enough to smooth
out the inherent ordering in the Li-Sn-D slabs.
In order to obtain a thorough understanding of the segregation effect and the density profiles
in the Li-Sn slab systems, substantially larger systems with
much longer trajectories in MD simulations are needed in future works.
However, we consider the FPMD trajectories are still meaningful to yield important
properties of D, such as the formation and evaporation of D$_2$ molecules and LiD molecules at the Li-Sn surfaces.}

\subsection{Formation of D$_2$ Molecules}

\begin{figure}[h]
  \centering
  \includegraphics[width=0.8\textwidth]{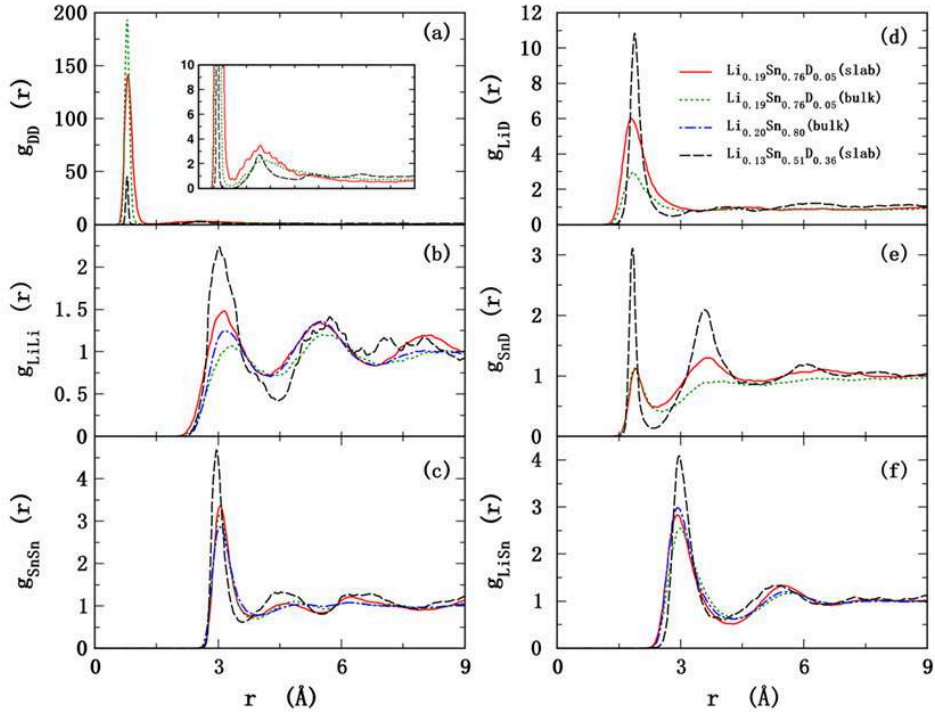}\\
\caption{(Color online) Partial pair distribution functions of
(a) D-D, (b) Li-Li, (c) Sn-Sn, (d) Li-D, (e) Sn-D, (f) Li-Sn
in the Li$_{0.19}$Sn$_{0.76}$D$_{0.05}$ slab (red lines),
the Li$_{0.19}$Sn$_{0.76}$D$_{0.05}$ bulk (green lines),
the Li$_{0.20}$Sn$_{0.80}$ bulk (blue lines),
\MC{and the Li$_{0.13}$Sn$_{0.51}$D$_{0.36}$ slab (black lines)} at 673 K.}
 %Partial pair distribution functions of
%(a) D-D, (b) Li-Li, (c) Sn-Sn, (d) Li-D, (e) Sn-D, (f) Li-Sn
%in the simulated liquid Li$_{0.19}$Sn$_{0.76}$D$_{0.05}$ slab at 673 K.}
\label{PPDF}
\end{figure}

Figure~\ref{PPDF} illustrates the partial pair distribution functions $g_{\alpha\beta}(r)$ of D-D, Li-Li, Sn-Sn, Li-D, Sn-D, and Li-Sn, which were obtained with FPMD simulations in the liquid Li$_{0.19}$Sn$_{0.76}$D$_{0.05}$
\MC{and Li$_{0.13}$Sn$_{0.56}$D$_{0.36}$ slabs} at 673 K.
As comparisons, we also show the pair distribution functions for bulk Li$_{0.20}$Sn$_{0.80}$ and  Li$_{0.19}$Sn$_{0.76}$D$_{0.05}$ systems.

The partial pair distribution function $g_{\alpha\beta}(r)$ with
two different atomic species $\alpha$ and $\beta$ is defined as
\begin{equation}
g_{\alpha\beta}(r)=\frac{N_{\alpha}+N_{\beta}}{\rho N_{\alpha}N_{\beta}}
\sum_{i=1}^{N_{\alpha}}\sum_{j=1}^{N_{\beta}}\langle\delta(\mathbf{r+R_{i}-R_{j}})\rangle,
\end{equation}
where $\rho$ is the atomic density,
$N_{\alpha}$ and $N_{\beta}$ are the numbers of atoms for species $\alpha$ and $\beta$, respectively.
$\mathbf{R_{i}}$ and  $\mathbf{R_{j}}$ are the atomic positions of atoms $i$ and $j$.
\MC{The above formula can be directly applied to bulk systems. For a slab system, we only
compute $g_{\alpha\beta}(r)$ for the bulk region of the slab, which has been defined above. }
First of all, we can see that the partial pair distribution functions
in Figures~\ref{PPDF}(b-f) all exhibit liquid-like structural characteristics.
As can be seen from Figure~\ref{PPDF}(a), $g_{DD}(r)$ with peaks centered at around 0.8 \AA~suggests
that D$_2$ molecules form in the liquid Li$_{0.19}$Sn$_{0.76}$D$_{0.05}$ slab and bulk,
as the typical bond length of a D$_2$ molecule is 0.74 \AA.
Both in the bulk and in the slab, the Li-D pair distribution functions have strong peaks around 1.85\AA,
suggesting that Li-D may form strong bonds.
\MC{We can see that most partial pair distribution functions from the high-concentration slab system
exhibit more prominent peaks. The result indicates that the Li$_{0.13}$Sn$_{0.51}$D$_{0.36}$ slab system
is more structured as compared to the Li$_{0.19}$Sn$_{0.76}$D$_{0.05}$ slab system,
which may be caused by the presence of high-concentration D atoms in the system.
Besides, the result also implies that Li and Sn atoms in the high-concentration slab system
interact more strongly with D atoms than those in the low-concentration slab system.}

\begin{figure}
  \centering
  \includegraphics[width=0.6\textwidth]{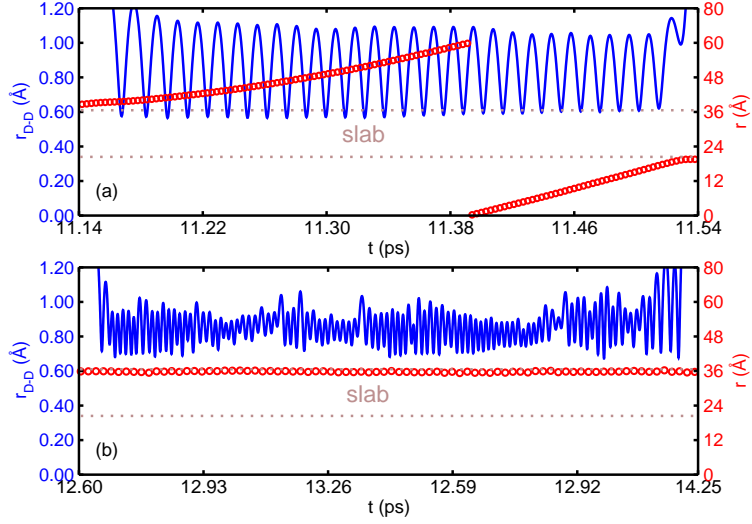}\\
  \caption{(Color online)
   \MC{Two stable formation events of D$_2$ molecules in the Li$_{0.19}$Sn$_{0.76}$D$_{0.05}$ slab system.
   The cell length along the surface normal direction $r$ is 60 \AA.
   For convenience, we plot $r=$60 to 80~\AA~ in order to see the oscillations of
   the bond length $r_{\mathrm{D-D}}$ of a D$_2$ molecule. Periodic boundary conditions are used.}
  }\label{fig:events}
\end{figure}

\begin{figure}
  \centering
  \includegraphics[width=0.8\textwidth]{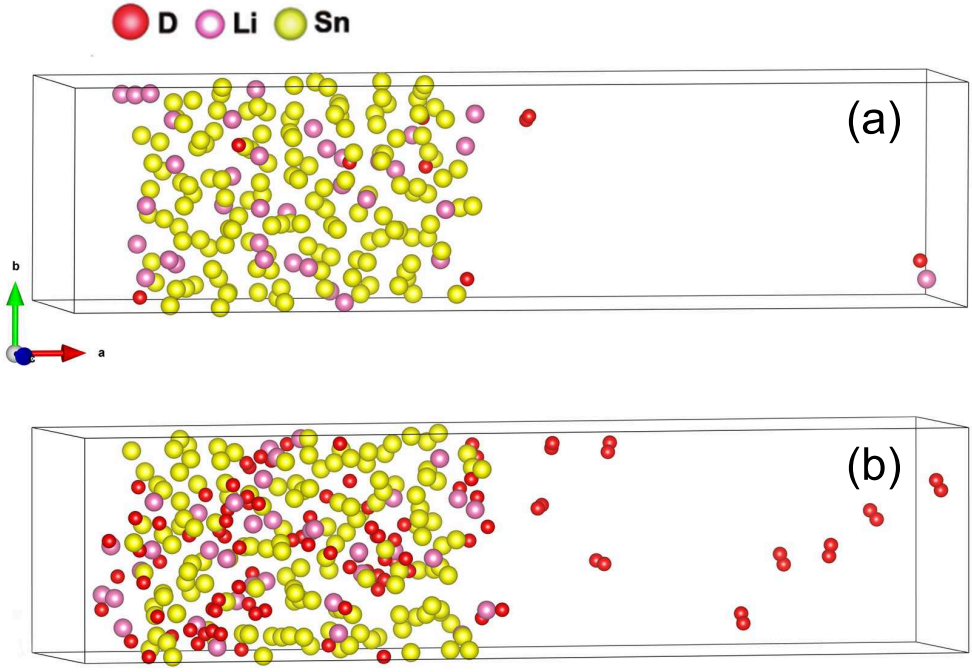}\\
  \caption{\MC{
  (Color online) Representative snapshots of (a) Li$_{0.19}$Sn$_{0.76}$D$_{0.05}$ slab system at 873 K
  and (b) Li$_{0.13}$Sn$_{0.51}$D$_{0.36}$ slab system at 673 K from first-principles
  molecular dynamics. The Li, Sn and D atoms are labeled with pink, yellow and red colors, respectively.
  }
  }\label{fig:stru}
\end{figure}

\begin{figure}
  \centering
  \includegraphics[width=0.6\textwidth]{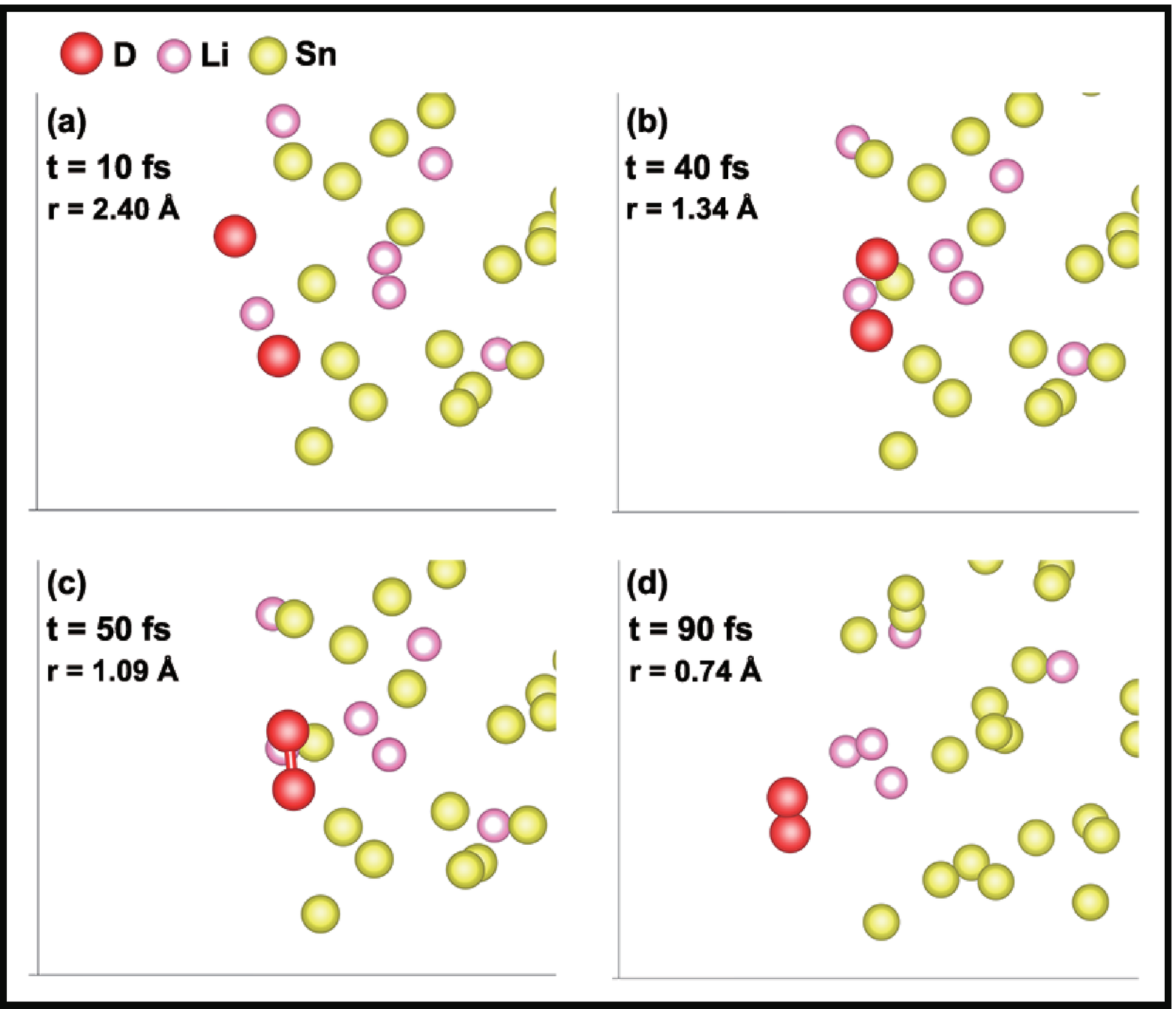}\\
  \caption{(Color online)
   A series of steps to form D$_2$ molecules in the Li$_{0.19}$Sn$_{0.76}$D$_{0.05}$ slab at 673 K.
   (a) Li and Sn trap a D atom at the surface. (b) Two D atoms come closer at the surface.
   (c) Formation of a D$_2$ molecule. (d) The D$_2$ molecule diffuses away from Li and Sn atoms.
   $t$ is the simulation time, whereas $r$ is the distance between the two D atoms.
  }\label{fig:snapshots}
\end{figure}

\MC{In the analyzed 20-ps FPMD trajectories of
the Li$_{0.19}$Sn$_{0.76}$D$_{0.05}$ slab,
we observe 6, 8, and 6 stable formation events of D$_2$ molecules at 573, 673, and 873 K, respectively.
Among the formation events, 4, 3, and 5 D$_2$ molecules respectively evaporate into the vacuum
in the system at 573, 673, and 873 K, whereas others only diffuse in the bulk region.
We show two examples in the Li$_{0.19}$Sn$_{0.76}$D$_{0.05}$ slab system in Figure~\ref{fig:events},
where we record the period from the formation of a D$_2$ molecule to the decomposition of a D$_2$ molecules into two D atoms.
From this figure, we observe oscillations of the bond length of two D$_2$ molecules;
one in the vacuum and the other in the slab near the surface. In Figure~\ref{fig:events}(a),
a D$_2$ molecule evaporates into the vacuum at around 11.14 ps and then redeposits into another surface of the slab at about 11.54 ps.
In Figure~\ref{fig:events}(b), a D$_2$ molecule forms at the surface at about 12.60 ps and decomposes to D atoms at around 14.25 ps.
Due to the periodic boundary conditions, most of the D$_2$ molecules that evaporate into the vacuum
redeposit into the other surface of the slab.
The number of D$_2$ molecules observed in the vacuum is therefore
smaller than the number of evaporation events, as shown in Figure~\ref{fig:stru}(a).
In contrast, we find several D$_2$ molecules that exist in the vacuum
of the Li$_{0.13}$Sn$_{0.51}$D$_{0.36}$ system, and D clusters only exist in the
bulk region, as illustrated in Figure~\ref{fig:stru}(b).
This can be explained as the D atoms are saturated in the high-concentration slab system,
so more D$_2$ molecules evaporate to the vacuum.
Furthermore, we do not observe evaporation
of single D atoms or D clusters in the high-concentration case.
}

During the preparation of Li-Sn-D samples, each D atom was randomly
implanted in the liquid Li-Sn slab before the simulations.
Hence, D$_2$ molecules can only form from individual D atoms in the Li-Sn sample
during the simulations,~\cite{19NF-Beatriz} and then escape into the vacuum area of the simulation cell.
In addition, D$_2$ molecules are
only found to dissociate into D atoms inside the Li-Sn slab but not in the vacuum.
\MC{The evaporation of D$_2$ molecules instead of D atoms or D clusters in both low and high concentrations of
Li-Sn-D systems provides some clues
to the recent experiment.~\cite{17NME-Loureiro}
A very recent work has also found the formation of D$_2$ molecules
in bulk liquid Li-Sn systems with FPMD simulations.~\cite{19NF-Beatriz}
However, due to the limitations of the FPMD simulations that trajectories are still too short,}
it is very difficult to compare even quantitatively the calculated percentage of released D$_2$ molecules
with the experiments, where almost all D$_2$ molecules are released.~\cite{17NME-Loureiro}
Additionally, several factors that exist in experiments but are not considered
in our simulations. For example, in the experiments, the liquid slabs may not be in the equilibrium state due to
the continuous bombardments of deuterium atoms and the defects in the slab may also affect the final results.
\MC{
However, by investigating the 20 stable formation events of D$_2$ molecules from the low concentration case
at temperatures of 573, 673, and 873 K, we analyze the detailed forming processes of D$_2$ molecules.
We do not investigate the formation of D$_2$ molecules in the high-concentration case because
a large number of D atoms exist in the slab, some of which form D clusters. Therefore,
it is challenging to have an algorithm to define D$_2$ molecules in D clusters.
In addition, even if we can identify D$_2$ molecules in the high-concentration case,
it is still difficult to get a clear picture about how D atoms interact with Li and Sn atoms
before a D$_2$ molecule forms, because the existence of adjacent D atoms complicates the analysis.
In contrast, in the low-concentration case,
the generation of each D$_2$ molecule can be nonambiguously identified and the forming process can be analyzed.
More importantly, the low-concentration case is a more realistic case
in a fusion device that uses liquid metals as shields because huge amounts of D atoms
bombarding liquid wall is unlikely to occur in an operating fusion device.
}
%As we will show below, our results strongly suggest that the formation and releasing D$_2$ molecule in the Li-Sn liquid alloy,
%which may be an effective way for recycling of D atoms in liquid Li-Sn slabs.

\MC{In a previous FPMD simulation work of inserted D atoms in bulk Li}~\cite{16NF-Chen}, the formation of D$_2$ molecules
did not occur when the proportion of D atoms is smaller than that of Li atoms
due to the strong bonding between Li and D atoms.
Only a small amount of D$_2$ molecules were recorded
when the proportions of D and Li are equal.
\MC{In another FPMD simulation work,} no D$_2$ formation with a long lifetime was observed in pure liquid Sn
~\cite{17JCP-Liu}.
In stark contrast, in the Li-Sn slab,
a considerable number of D$_2$ molecules form in a system
with more Li than D,
indicating there may be a different mechanism to form D$_2$ in the Li-Sn eutectic.
Based on our analysis,
there are typically four steps to form D$_2$ molecules in the Li-Sn slab as shown in Figure~\ref{fig:snapshots}
(see the movie in the SM \cite{SM}):
In the first step, a D atom is first trapped by a Li atom in the Li-Sn slab to form a Li-D pair, as shown in Figure~\ref{fig:snapshots}(a).
The trapped D keeps moving at a relatively low velocity,
setting up a stable environment with adjacent Li and Sn atoms
and waiting for other D atoms to form a D$_2$ molecule.
Figure~\ref{PPDF}(a) further demonstrates that D atoms interact strongly with Li atoms,
as can be seen by the high first peak of $g_{LiD}$(r).
Next, Figure~\ref{fig:snapshots}(b) shows that a second D atom comes close to the first D,
while the position of the Li-D pair does not change much.
As will be explained later, we find that the second \MC{D atom slows down and diffuses} towards the first D.
Third, a D$_2$ molecule starts to form by triggering the charge transfer
from D to its adjacent Sn atoms ({\it vide infra}), as shown in Figure~\ref{fig:snapshots}(c).
In the last step, the newly formed D$_2$ molecule diffuses away from its Li and Sn neighbors
as illustrated in Figure~\ref{fig:snapshots}(d).
We also provide a movie~\cite{SM} that shows the formation of a D$_2$ molecule
near the surface of Li$_{0.19}$Sn$_{0.76}$D$_{0.05}$ system (673 K) and evaporation of this D$_2$ molecule into the vacuum
(see the Supporting Information).

\begin{figure}
  \centering
  \includegraphics[width=0.52\textwidth]{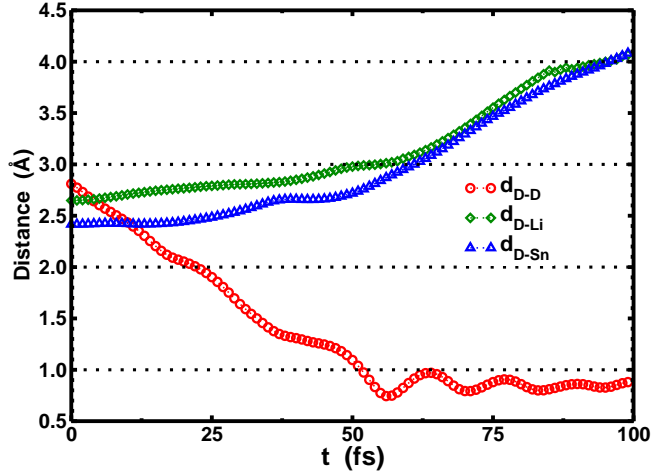}\\
  \caption{(Color online)  Distances from D atoms to its adjacent atoms in the Li$_{0.19}$Sn$_{0.76}$D$_{0.05}$ slab
  during the formation of D$_2$ molecules. $d_{\mathrm{D-D}}$, $d_{\mathrm{D-Li}}$, and $d_{\mathrm{D-Sn}}$ are defined as the
  distances between D and its adjacent D, Li, and Sn atoms, respectively.
  \MC{The results are averaged over 20 stable formation events of D$_2$ molecules found in the Li$_{0.19}$Sn$_{0.76}$D$_{0.05}$ slab
  at temperatures of 573, 673, and 873 K.}
  }\label{fig:properties}
\end{figure}

\begin{figure}
  \centering
  \includegraphics[width=0.52\textwidth]{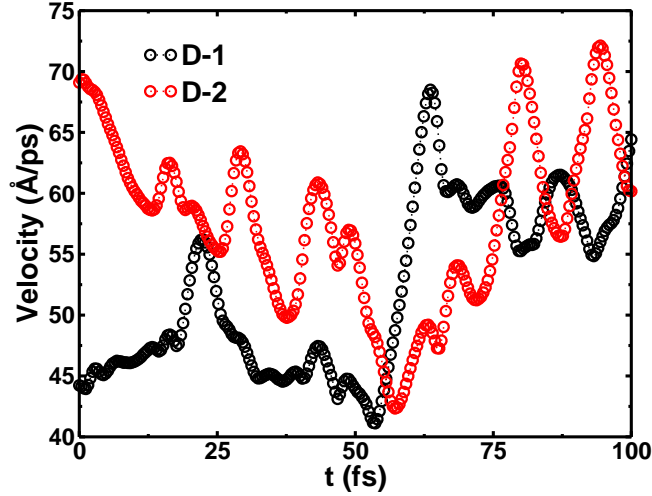}\\
  \caption{(Color online)
  Velocities of two D atoms (labeled as D-1 and D-2) during the formation of D$_2$ molecules in the Li$_{0.19}$Sn$_{0.76}$D$_{0.05}$ slab.
  \MC{The results are averaged over 20 stable formation events of D$_2$ molecules found in the Li$_{0.19}$Sn$_{0.76}$D$_{0.05}$ slab
  at temperatures of 573, 673, and 873 K.}
  }\label{fig:properties2}
\end{figure}

\begin{figure}
  \centering
  \includegraphics[width=0.52\textwidth]{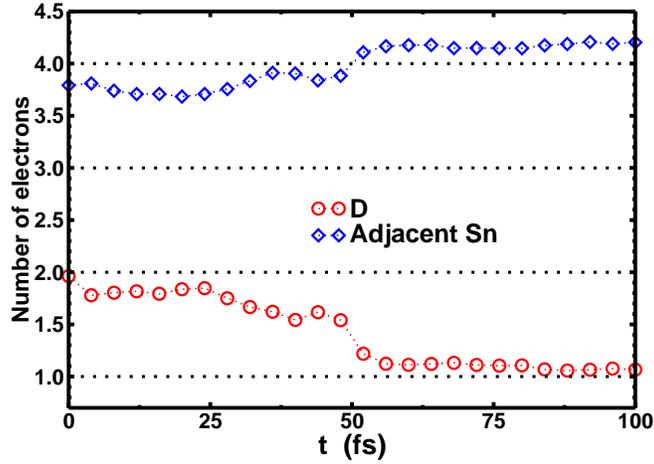}\\
  \caption{(Color online)
  Number of electrons on D and its adjacent Sn atoms during the formation of a D$_2$ molecule in the Li$_{0.19}$Sn$_{0.76}$D$_{0.05}$ slab.
 \MC{The results are averaged over 20 stable formation events of D$_2$ molecules found in the Li$_{0.19}$Sn$_{0.76}$D$_{0.05}$ slab
  at temperatures of 573, 673, and 873 K.}
  }\label{fig:properties3}
\end{figure}

\subsection{Processes of Forming D$_2$}
We systematically analyze the processes of forming D$_2$ molecules in order to quantify the above observations in simulations.
\MC{For those properties illustrated in Figure~\ref{fig:properties},
Figure~\ref{fig:properties2}, and Figure~\ref{fig:properties3},
we average 20 stable formation events of D$_2$ molecules.
The 20 events are collected from the 6, 8, and 6 stable formation events of D$_2$ in the low-concentration system at
temperatures of 573, 673, and 873 K, respectively.
Although more stable formation events of D$_2$ molecules are expected at higher temperatures,
it is counter intuitive that only 6 events of D$_2$ are found at 873 k, which is smaller than the 8
events of D$_2$ found at 673 K.
Nevertheless, at 873 K, we record 10 evaporation events of LiD molecules ({\it vide infra}),
indicating that some of the D atoms in the low-concentration slab system are involved in forming LiD molecules. }
In particular, we track the D-D pairs in the low-concentration Li-Sn slab
and report their interesting dynamics.
The criterion to track D-D pairs was selected when any two D atoms
come close within 1.1~\AA~and last for more than 0.2 ps.
Moreover, we chose a time window of 100 fs that is relevant to the formation of D$_2$ molecules,
emphasizing the formation moment which occurs at around 50 fs
corresponding to Figures~\ref{fig:snapshots}(a-c).
By investigating the changes of properties before and after the formation of D$_2$ molecules within the 100 fs,
we can obtain insights into the fundamental process of forming D$_2$ molecules.
Specifically, the analyzed time-dependent properties include distances of D to its neighbors,
instantaneous velocities of D atoms, and number of electrons on D,
which are respectively shown in Figure~\ref{fig:properties},
Figure~\ref{fig:properties2}, and Figure~\ref{fig:properties3}, respectively.
\MC{We point out that due to the limited size of systems utilized in our FPMD simulations,
the structural and dynamical properties of atoms in the slab could be affected.
However, in our analysis of these time-dependent properties, we only focus on a local environment
around the two D atoms that form a D$_2$ molecule. Therefore, we expect
that the size effects are not substantial in affecting these properties.
In addition, the number of D atoms in the low-concentration system is only 10 due to the
limited size of the system, we observe 20 stable formation events of D$_2$ molecules from 573, 673, and 873 K.
Although the current work could not provide more detailed temperature-dependent forming mechanisms for D$_2$ molecules,
we still consider the averaged properties from the 20 events are sufficient to provide useful information.}

Figure~\ref{fig:properties} shows that the distance between two D atoms ($d_{\rm D-D}$)
quickly shortens approximately from 2.8 to 0.8 \AA~ in the first 50 fs before the formation of D$_2$,
and then stably oscillates around 0.8 \AA~ with a damped amplitude up to 100 fs.
The value of 0.8 \AA~ is the location of the first peak of the partial pair distribution function g$_{\rm DD}$(r),
as illustrated in Figure~\ref{PPDF}(a).
Meanwhile, the distances between D atoms and its adjacent atoms, i.e., $d_{\rm D-Li}$ and $d_{\rm D-Sn}$, slowly
elongates from around 2.5 to 2.8 \AA~during the first 50 fs,
which can be understood since Li and Sn atoms should provide spaces for two neighboring D atoms to meet
and the process of forming D$_2$ starts.
Meanwhile, as displayed in Figures~\ref{PPDF}(d) and (e), D atoms start to deviate from positions of the first peak in
g$_{\rm LiD}$(r) and g$_{\rm SnD}$(r), indicating that the interactions between D and Li/Sn atoms are weakened.
We then see a dramatic increase of $d_{\rm D-Li}$ and $d_{\rm D-Sn}$ from around 2.8 to 4.0 \AA~within the second 50 fs,
suggesting that the D$_2$ molecule quickly diffuse away after its formation (see Figure~\ref{fig:snapshots}(d) and the movie in the Supplementary Materials~\cite{SM}).

The \MC{averaged} velocities of the selected D-D pairs
during the formation of D$_2$ molecules are illustrated in Figure~\ref{fig:properties2}.
\MC{One of the two D atoms (labeled as D-1) has a velocity
that is slower than the velocity of the other D atom (labeled as D-2) in the first 50 fs,}
indicating that the two D atoms are experiencing different environments, i.e.,
the D-1 atom is trapped by surrounding Li and Sn atoms and the D-2 atom diffuses
freely in the liquid with its velocity abruptly drops.
\MC{
During the formation of D$_2$ molecule at around 50 fs,
the two D atoms slow down and we know from Figure~\ref{fig:properties} that a bond forms between them.
Next, after the formation of a D$_2$ molecule,
both D-1 and D-2 atoms experience a vast increase of velocity after the 50 ps,
and then diffuse together with similar velocities.}

In Figure~\ref{fig:properties3},
we also compute the averaged charge change of D atoms and their adjacent atoms during the
formation of D$_2$ molecules. We utilized the  Bader charge analysis method~\cite{09JPCM-Tang}.
% was adopted to compute the number of electrons around each atom.
From this analysis, we find electrons transfer from D to Sn atoms during the formation of D$_2$ molecules.
Quantitatively, before forming D$_2$ molecules at around 50 fs, more than 1.5 electrons
are assigned to a D atom while less than 4.0 valence electrons per atom are on its nearby Sn atoms.
In the first 50 fs, a small amount of electrons (less than 0.5 electrons per atom) on D transfer to Sn.
When D$_2$ forms at around 50 fs, the number of electrons on the D atom abruptly drops to about 1.0 within a very short time,
which occurs at the same time when the changes of distances and velocities of D atoms occur.
Meanwhile, the electrons on Sn are accumulated to above 4.0.
The resulting number of electrons on D is slightly above 1.0,
which is close to that of a single D atom.
Furthermore, the D-D pair therefore has a bond of around 0.8 {\AA}
with two electrons.
Therefore, we identify the formation of D$_2$ molecules in the liquid Li-Sn slab based on all of the above considerations.
It should be noted that no electrons are assigned to Li according to the Bader analysis,
suggesting Li loses all its $2s$ valence electrons in the metallic liquid Li-Sn slab.

\begin{figure}
  \centering
  \includegraphics[width=0.95\textwidth]{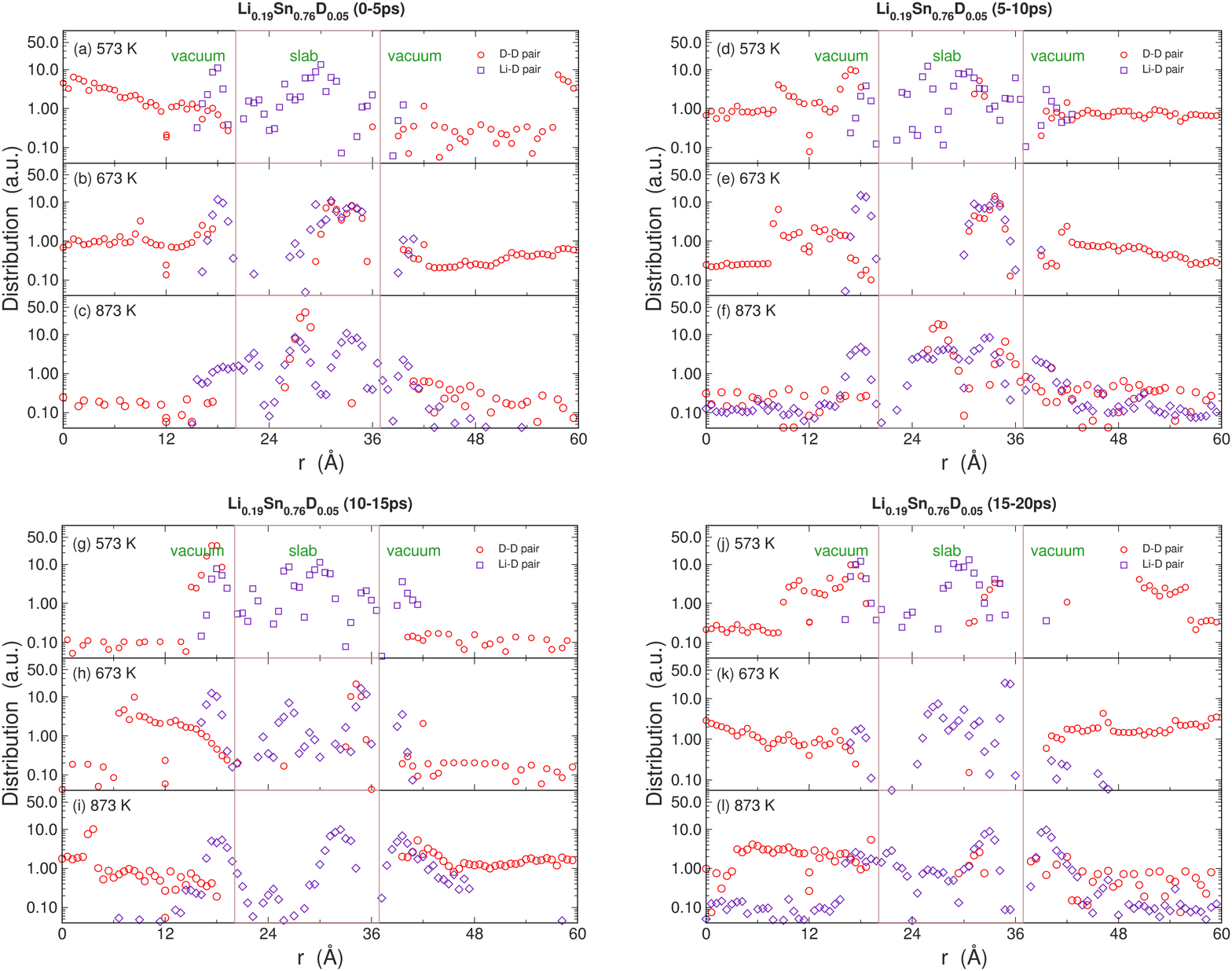}\\
  \caption{(Color online) Distributions of D-D and Li-D pairs
  in the Li$_{0.19}$Sn$_{0.76}$D$_{0.05}$ slab system along the surface normal direction
  at 573, 673 and 873 K. \MC{Distributions of D-D and Li-D pairs are divided into four
  parts according to the length of trajectories, i.e., (a-c) 0-5, (d-f) 5-10, (g-i) 10-15, (j-l) 15-20 ps.}
  The region within the two solid brown lines is defined as the bulk of Li-Sn.
 }\label{fig:LiD}
\end{figure}

\subsection{Formation of LiD Molecules}

The FPMD simulations show that the D$_2$ molecules diffuse rapidly in the
Li-Sn eutectic and some of them evaporate into the vacuum.
\MC{Although we observe no evaporation of single D atoms or Li$_2$ molecules
in liquid Li-Sn slab systems from the FPMD trajectories,}
we find the existence of LiD molecules in vacuum,
which is not surprising since Li and D atoms form strong ionic bonds \MC{and tend to stay near the surface}.
\MC{To be specific, we observe 10 evaporation events for LiD molecules in the low-concentration slab system at 873 K,
while only 1 evaporation event of LiD molecules is found in the high-concentration slab system at 673 K.}
\MC{Additionally, according to the clues provided by the atomic decomposed density profiles illustrated in Figure~\ref{fig:decompose},
we respectively observe 2 and 1 events for a single Li atom that appears in the vacuum region
of the low- and high-concentration slab systems at 873 K.
Based on the above findings, we focus on discussing the formation of LiD molecules
in the low-concentration slab system at 873 K since more evaporation cases are observed in this system
than those in other systems.}

Because Li could potentially contaminate the plasma,
it is important to know how the LiD molecules evaporate compared to D$_2$.
Figure~\ref{fig:LiD} depicts the distributions of D$_2$ and LiD molecules in
the low-concentration Li-Sn slab at 573, 673, and 873 K.
\MC{We divide the 20 ps trajectory into four parts, i.e., 0-5, 5-10, 10-15, 15-20 ps.}
D$_2$ molecules distribute in both liquid Li-Sn slab and vacuum at all temperatures considered.
In contrast, Li-D pairs mostly reside in the Li-Sn slab at 573 and 673 K,
and are occasionally recorded around the surface.
At the higher temperature of 873 K, we find a small amount of LiD molecules
escape to the vacuum, although the chance to find the LiD molecules
is almost an order of magnitude smaller than that of D$_2$ molecules.
This finding is significant, because it suggests that the facility should work in an optimal temperature window below
873 K, which allows efficient recycling of D atoms but reduce the risk of LiD molecules from contaminating the plasma.

\section{Conclusions}
To summarize, we carry out FPMD simulations to investigate
the retention and recycling of deuterium in liquid Li-Sn
with two different concentrations (Li$_{0.19}$Sn$_{0.76}$D$_{0.05}$ and Li$_{0.13}$Sn$_{0.51}$D$_{0.36}$)
at three temperatures of 573, 673, and 873 K.
First of all, we find the formation of D$_2$ molecules in all of the simulations.
Importantly, we unveil a cooperative mechanism of forming D$_2$ molecules in liquid Li-Sn,
where Li atoms behave as catalytic centers, allowing efficient generation
of D$_2$ molecules in liquid metal Li-Sn.
During the formation process, Li traps a D atoms and another D atom comes close to
trigger the charge transfer from D atoms to their nearby Sn atoms.
After formation, the D$_2$ molecules diffuse promptly in the Li-Sn liquid and evaporate to the vacuum.
The above mechanism reduces the retention of D in liquid Li-Sn slabs, and
is beneficial for the recycling of D atoms.
This work therefore provides new clues to a recent experimental finding~\cite{17NME-Loureiro}
that Li-Sn eutectic has an unexpected low retention rate of hydrogen isotopes.
Furthermore, although we do not observe evaporation of single D atoms or
Li$_2$ molecules in our simulations,
\MC{we occasionally observe the evaporation of LiD molecules in the high-concentration slab system at 673 K
and the appearance of Li atoms in the vacuum in both concentrations of systems at 873 K. Importantly,}
we observe \MC{a substantial amount of} LiD molecules,
\MC{which form and evaporate in the low-concentration slab system} at 873 K.
We predict a temperature window below 873 K, allowing
efficient recycling of D atoms but reduce the risk of LiD molecules from contaminating the plasma.
The new findings deepen our understanding of interactions between hydrogen isotopes and liquid Li-Sn
and can be potentially used for improving the design of magnetically confined fusion reactors.
Because the FPMD simulations of liquid slabs are computationally very expensive, currently we can
only simulate a slab with a few hundred atoms for short period of times, which unavoidably suffers from size effects.
Phenomena such as segregation of Li atoms to the liquid surface~\cite{17FED}
cannot be fully reproduced in the FPMD simulations.~\cite{19NME-Beatriz} Simulations on much larger systems and for a longer simulation time using reliable potentials constructed by machine learning~\cite{18L-Zhang}
is a promising routine along this direction.

\section{Data availability statement}
The data that support the findings of this study are available from the corresponding author upon reasonable request.

\acknowledgements
Daye Zheng and Zhen-Xiong Shen contributed equally to this work.
This work was funded by the National Key Research and Development Program of China (Grants No. 2016YFB0201202), and
the Chinese National Science Foundation Grant number 11774327.
XR acknowledges the support of the Chinese National Science Foundation Grant No. 11574283 and 11874335.
The numerical calculations have been done on the USTC HPC facilities.
Part of the numerical simulations was performed on the High Performance Computing Platform of CAPT.

%\bibliographystyle{apsrev}
%\bibliography{LiSnD_ref}

\end{document}